\documentclass[sigconf,nonacm]{acmart}   %

\newcommand\name{\textsc{TokenMinds}}
\usepackage[utf8]{inputenc} %
\usepackage[T1]{fontenc}    %
\usepackage{hyperref}       %
\usepackage{url}            %
\usepackage{booktabs}       %
\usepackage{amsfonts}       %
\usepackage{nicefrac}       %
\usepackage{microtype}      %
\usepackage{xcolor}         %
\usepackage{makecell}
\usepackage{graphicx}
\usepackage[]{algorithmic} 
\usepackage{booktabs}
\usepackage{multirow}
\usepackage[ruled,linesnumbered]{algorithm2e}

\makeatletter
\def\@affiliationfont{\normalsize\normalfont} %
\makeatother
\begin{document}
\title{\name{}: Pretrained User Tokens and Embeddings for User Understanding in Large Recommender Systems} 

\author{Qingyun Liu}
\email{qyl@google.com}
\affiliation{%
  \institution{Google DeepMind}
  \country{}}

\author{Bo Yan}
\email{boyanby@google.com}
\affiliation{%
  \institution{YouTube}
  \country{}}

\author{Yang Liu}
\email{ylyangliu@google.com}
\affiliation{%
  \institution{YouTube}
  \country{}}

\author{Yuji Roh}
\email{yujiroh@google.com}
\affiliation{%
  \institution{Google DeepMind}
  \country{}}

\author{Ekansh Sharma}
\email{ekansharma@google.com}
\affiliation{%
  \institution{YouTube}
  \country{}}

\author{Likang Yin}
\email{lkyin@google.com}
\affiliation{%
  \institution{YouTube}
  \country{}}

\author{Emma Olowo}
\email{eemasealu@google.com}
\affiliation{%
  \institution{YouTube}
  \country{}}

\author{Yuxuan Li}
\email{yuxuanli@google.com}
\affiliation{%
  \institution{YouTube}
  \country{}}

\author{Yikun Jiang}
\email{yikunj@google.com}
\affiliation{%
  \institution{YouTube}
  \country{}}

\author{Diego Uribe}
\email{diegouribe@google.com}
\affiliation{%
  \institution{YouTube}
  \country{}}

\author{Saksham Aggarwal}
\email{skshm@google.com}
\affiliation{%
  \institution{YouTube}
  \country{}}

\author{Siqi Wu}
\email{siqiwu@google.com}
\affiliation{%
  \institution{YouTube}
  \country{}}

\author{Yuan Hao}
\email{yuanhao@google.com}
\affiliation{%
  \institution{Google DeepMind}
  \country{}}

\author{Min-hsuan Tsai}
\email{mhtsai@google.com}
\affiliation{%
  \institution{Google DeepMind}
  \country{}}

\author{Vikas Kedigehalli}
\email{vikasrk@google.com}
\affiliation{%
  \institution{YouTube}
  \country{}}

\author{Lukasz Heldt}
\email{heldt@google.com}
\affiliation{%
  \institution{YouTube}
  \country{}}

\author{Lichan Hong}
\email{lichan@google.com}
\affiliation{%
  \institution{Google DeepMind}
  \country{}}

\author{Li Wei}
\email{liwei@google.com}
\affiliation{%
  \institution{YouTube}
  \country{}}

\author{Xinyang Yi}
\email{xinyang@google.com}
\affiliation{%
  \institution{Google DeepMind}
  \country{}}

\renewcommand{\shortauthors}{Liu et al.}

\begin{abstract}
User modeling in industrial recommender systems typically produces dense embeddings, which suffer from representational constraints due to fixed-dimensional vectors.
As an emerging alternative for discrete user representation, LLM-generated natural-language user profiles often capture language semantics rather than sequential patterns in user behaviors.
Meanwhile, Semantic ID (SID) is effective for item representation, yet existing discrete user representations remain largely confined to quantizing already-pooled user vectors, clustering historical watches, or predicting coarse topic clusters solely for candidate retrieval. We propose \name{}, an industrial system that extends the PLUM framework~\cite{he2025plum} from SID-based item retrieval to generative sequential user modeling. Built on an encoder-decoder architecture adapted from pretrained LLMs, \name{} introduces a dual-output design: the encoder outputs \emph{a dense user embedding} capturing a holistic user state, while the decoder generates \emph{discrete SID-based user tokens} representing multi-faceted future interests. Grounded in item semantics, these tokens preserve distinct interests that a single pooled vector would average away. Additionally, a shared SID vocabulary across long- and short-form videos unifies cross-scenario behaviors into a single model, cutting upstream training and serving compute by 50\% and 31\%. Furthermore, leveraging the text vocabulary inherited from the pretrained LLM, \name{} interleaves textual search queries directly with watch histories to capture explicit user intent. We validate \name{} through offline evaluations and live A/B experiments across YouTube's primary ranking surfaces, serving billions of users asynchronously: SID-based user tokens are effective on their own, yield additive gains alongside dense embeddings, and together outperform the previous-generation, non-LLM user model in production.

\end{abstract}

\maketitle

\section{Introduction}
\label{sec:intro}

Sequential user modeling is central to modern recommender systems, leveraging interaction histories to drive personalization~\cite{lyu2025dv365,chai2025longer,chang2023latent}. Today, Large Embedding Models (LEMs)~\cite{covington2016deep,coleman2023unified} dominate industrial practice by using massive tables to represent high-cardinality item IDs. While effective at memorizing user--item interactions, LEMs fundamentally struggle with generalization~\cite{singh2024better}.

Conversely, Large Language Models (LLMs) shift parameter budgets from massive embedding tables to neural networks, showing strong potential in feature extraction, LEM integration~\cite{zhou2025hymirec}, and end-to-end sequence modeling~\cite{zhou2025onerec,zhou2025onerecv2,liu2025onerecthink}. However, applying LLMs to industrial recommendation faces two challenges: a modality gap when processing massive, non-textual ID spaces ({\em e.g.}, billions of videos), and inherited limitations when coupled with traditional ID tables, such as scaling bottlenecks, vocabulary churn, and expressiveness limits~\cite{singh2024better,weller2026theoretical}. On the \emph{item} side, Semantic IDs (SIDs)~\cite{rajput2023TIGER} overcome table dependencies by tokenizing items into hierarchical codewords from content semantics~\cite{singh2024better}, and the PLUM framework~\cite{he2025plum} bridges the modality gap by aligning the SID vocabulary with pretrained LLMs via Continued Pre-Training (CPT) for generative item retrieval.

However, how to construct expressive, general-purpose \emph{user} representations remains an open challenge. The predominant industrial user representation is a dense embedding~\cite{pancha2022pinnerformer,lyu2025dv365} that compresses a user's full spectrum of interests into one fixed-dimensional vector and may lose fine-grained signals; multi-interest variants spread interests over several vectors~\cite{li2019mind,cen2020comirec} but are prone to interest collapse~\cite{wu2025gemirec}. As a discrete alternative, LLM-generated natural-language user profiles~\cite{wang2026genie,fazelnia2026textual} tend to capture language semantics rather than sequential behavior dynamics~\cite{kim2025lost}. Meanwhile, SID- or cluster-based user representations either quantize already-pooled user vectors into standalone user codebooks~\cite{qu2025tokenrec,lu2025userip}, structure already-consumed history through SID-guided sparse compression~\cite{yuan2026hisac} or interest clusters~\cite{lin2026uitp}, or decode discrete interests strictly for candidate retrieval~\cite{wang2024llms,wu2025gemirec}, rather than generatively predicting future user interests as general-purpose signals for downstream recommenders.

To address these challenges, we present \name{}, an industrial-scale user foundation model that extends PLUM~\cite{he2025plum} from item retrieval to generative sequential user modeling. Built on an LLM-adapted encoder-decoder architecture~\cite{rajput2023TIGER,zhou2025onerec}, \name{} introduces a dual-output design (Sec.~\ref{sec:enc_dec}): the encoder produces a dense user embedding capturing a holistic user state, while the decoder fully cross-attends over the unpooled history to generate discrete SID-based user tokens that capture multi-faceted future interests grounded in item semantics. Using production rankers as our primary evaluation testbed, we show that adapting discrete user tokens via co-trained sub-word embedding tables for user-side pooling or candidate-aware target attention, unlocks standalone gains and compounds additively with dense embeddings. Together, this dual representation outperforms the production LEM user model: it delivers gains on top of LEM while leaving LEM no residual value on primary rankers (Sec.~\ref{sec:exp_live}). To serve billions of users under strict latency budgets, \name{} runs atop an asynchronous caching infrastructure~\cite{li2024ubs} that decouples generative inference from real-time scoring (Sec.~\ref{sec:infra}).

Beyond single-scenario user modeling, because \name{}
operates in a shared token space inherited from a pre-trained LLM,
it naturally enables two further extensions (Sec.~\ref{sec:unified}). First, a shared SID vocabulary bridges long-form (LFV) and short-form videos (SFV) despite their disjoint ID spaces and distinct consumption patterns~\cite{liu2023multitask}. By conditioning on format tokens (\texttt{LFV}/\texttt{SFV}) during training and applying \textit{multi-context decoding} at inference, a single model encodes the chronologically interleaved history once and decodes format-specific user tokens in parallel without multi-model overhead. Second, because the vocabulary retains the text tokens of the pretrained LLM, textual search queries can be integrated directly into the watch sequence to capture explicit user intent.

We summarize our contributions below:
\begin{itemize}
    \item \textbf{Dual-Output Generative User Model:} We propose \name{}, an industrial user foundation model that jointly generates \emph{dense continuous user embeddings} from its encoder and \emph{discrete SID-based user tokens} from its decoder.
    \item \textbf{Tokenized User Adaptation \& Outperforming the LEM User Model:} We show that co-trained learnable embeddings unlock the effectiveness of discrete user tokens in production rankers, and that combining tokens with dense embeddings yields additive gains. Together, \name{}'s dual representation outperforms the production LEM user model: it adds gains on top of LEM across surfaces while rendering it redundant.
    \item \textbf{Cross-Scenario Unification \& Search-Query Integration:} By unifying LFV and SFV training and introducing multi-context decoding, \name{} cuts upstream training/serving cost by 50\%/31\% while boosting fresh-content engagement. Furthermore, interleaving textual search queries into the unified sequence yields additional live gains.
    \item \textbf{Billion-User Production Deployment:} \name{} has been deployed on $100\%$ of global YouTube traffic for over nine months, serving both LFV and SFV recommendation across multiple ranking models and retrieval.
\end{itemize}

\section{Related Work}
\label{sec:related_work}

\paragraph{Sequential User Modeling.}
Sequential recommendation has progressed from early deep models~\cite{covington2016deep, hidasi2016rnn} to self-attentive Transformers~\cite{kang2018sasrec} or trillion-parameter generative transducers~\cite{zhai2024hstu}, with recent industrial systems scaling to ultra-long histories~\cite{si2024twinv2,lyu2025dv365, chai2025longer, guan2026make}. 
To amortize this cost across surfaces, industrial systems pre-compute a user embedding upstream and serve it to downstream models~\cite{pancha2022pinnerformer, yang2023LURM, zhang2024sum, chen2025pinfm}.
The predominant user representation, {\em i.e.}, a single fixed-dimensional vector, creates an information bottleneck~\cite{weller2026theoretical, xiong2026llatte}.
Multi-interest models relax this with several vectors per user~\cite{li2019mind, cen2020comirec}, yet the vectors often suffer from interest collapse unless held apart by soft regularizers that trade off accuracy~\cite{wu2025gemirec}.

\paragraph{Discrete \& SID-Based User Representations.}
To move beyond continuous vectors, LLM-based recommenders distill user histories into natural-language personas~\cite{wang2026genie,fazelnia2026textual}, yet they often capture language semantics rather than sequential patterns in user behaviors~\cite{kim2025lost} and are misaligned with non-textual rankers~\cite{wang2024llm4msr,lu2025userip}. In parallel, hierarchical Semantic IDs (SIDs) learned via residual quantization~\cite{rajput2023TIGER} provide generalization and vocabulary stability for \emph{items}~\cite{singh2024better,he2025plum,chen2026tokenfactory}.
On the \emph{user} side, recent works quantize pooled user vectors or text-prompt embeddings into discrete codes~\cite{qu2025tokenrec,lu2025userip}, yet quantizing an already-compressed representation inherits its pre-quantization bottleneck.
Other production systems often structure \emph{already-consumed} history: using past item SIDs to sparsely route and aggregate historical item embeddings~\cite{yuan2026hisac} or grouping past watches into hierarchical interest sequences~\cite{lin2026uitp}. Additionally, other approaches generate discrete interests strictly for \emph{candidate retrieval} via $K{=}2$ coarse topic clusters~\cite{wang2024llms} or jointly trained interest codebooks~\cite{wu2025gemirec}.
In contrast, \name{} cross-attends over the full, unpooled history to autoregressively decode a user's \emph{future} interests as multiple SID-prefix trajectories in the item SID vocabulary, and serves these tokens alongside a dense embedding as general-purpose features across multiple downstream rankers.

\paragraph{Generative Recommendation.}
This paradigm reframes candidate retrieval and ranking as sequence-to-sequence generation over discrete item identifiers~\cite{rajput2023TIGER, he2025plum, zhou2025onerec}.
Unlike generative retrieval~\cite{he2025plum, zhou2025onerec} that predicts immediate next items, generative ranking~\cite{huang2025genrank} that scores candidates online, or task-aligned generative user models~\cite{zhang2025gpr}, \name{} decouples upstream user representation learning from specific downstream objectives. Specifically, \name{} extends the PLUM framework~\cite{he2025plum} to general-purpose upstream user modeling that aims to capture a wider spectrum of intents over a longer time window, avoiding the strict need to map back to specific, individual items. Concurrent works~\cite{yang2024liger,yang2026cobra} show that sparse SIDs and dense embeddings are complementary at the \emph{item} level for retrieval, while recent work~\cite{li2026tokens} studies downstream aggregation of tokenized items. \name{} extends this sparse--dense complementarity to the \emph{user} side, jointly serving SID-based user tokens and continuous embeddings to downstream models.

\paragraph{Cross-Scenario Modeling.}
Unifying user behaviors across heterogeneous surfaces is a core challenge in industrial recommenders~\cite{zang2022survey}.
Prior multi-scenario methods transfer multi-source histories via specialized ranking towers or adapters~\cite{ma2022mixed, yang2024mlora, wang2024llm4msr}, while recent industrial user foundation models project multi-surface interactions into a shared continuous representation~\cite{badrinath2025omnisage, he2026foundv2, xu2025redrec}.
Rather than relying solely on a shared continuous space, \name{} encodes interleaved cross-scenario histories into a unified intent embedding and scenario-specific discrete token representations, reducing upstream compute while preserving scenario-aware granularity.

\section{\name{} Framework}
We first overview the \name{} framework and detail the joint generation of discrete SID-based user tokens alongside dense embeddings. We then extend this architecture to cross-scenario modeling and search integration, discuss downstream token adaptation, and conclude with our asynchronous serving system.

\begin{figure}[t!]
    \centering
    \includegraphics[width=0.99\linewidth]{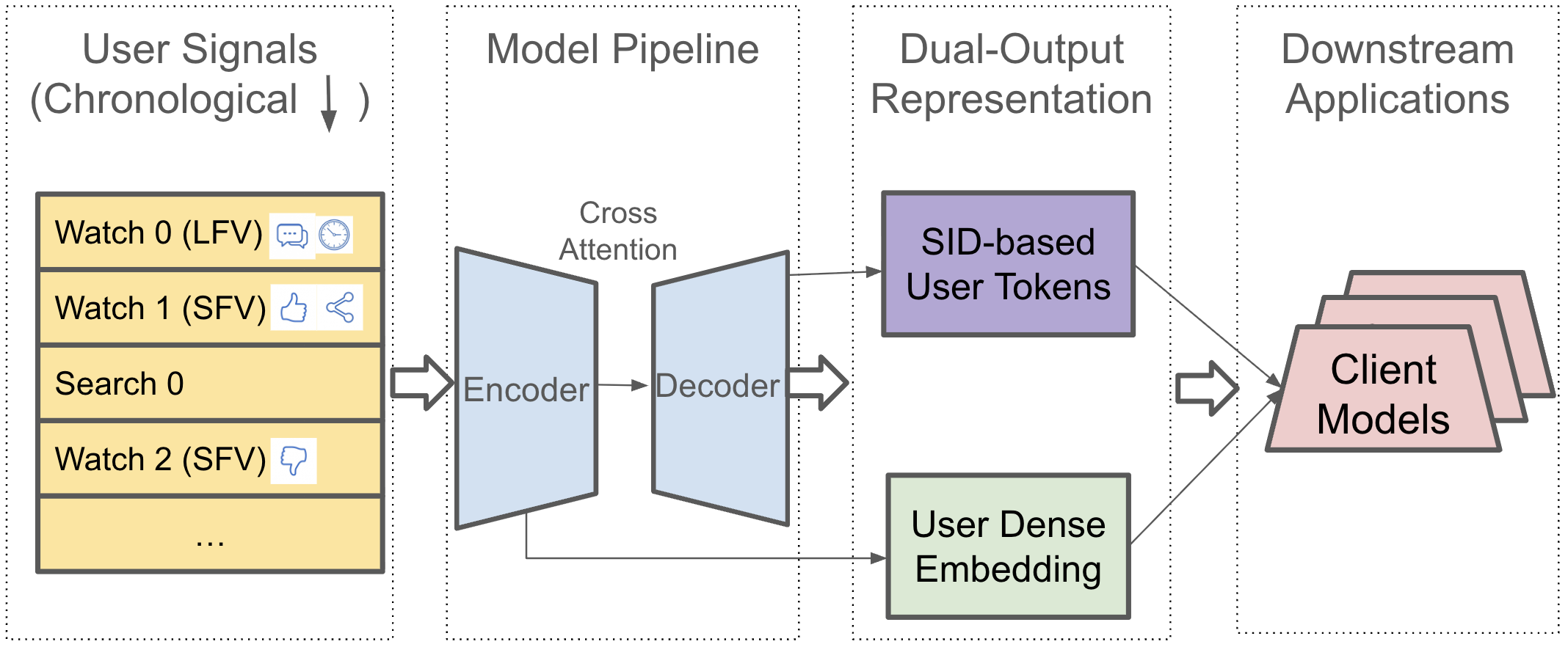}
    \caption{Overview of the \name{} framework. An encoder-decoder architecture processes heterogeneous user signals: watches across long- and short-form videos, search queries, and associated engagement features. It simultaneously produces dense user embeddings from the encoder and discrete SID-based user tokens from the decoder, which are served to downstream models.}
    \label{fig:framework}
\end{figure}

\subsection{System Overview}
\label{sec:overview}
We adopt a sequence-to-sequence (seq2seq) framework for \name{}, as illustrated in Fig.~\ref{fig:framework}. User behavior signals are tokenized into a sequential input and processed by a pretrained foundation model to produce both dense user embeddings and discrete SID-based user tokens. Since foundation models lack prior exposure to our video corpus, we follow PLUM~\cite{he2025plum} with Continued Pre-Training (CPT) to align the new SID modality with the model's existing knowledge (ablated in Sec.~\ref{sec:exp_offline}), followed by Supervised Fine-Tuning (SFT) for user modeling. We describe the inputs, representations, and architecture below.

\paragraph{User Data.} We incorporate disparate user behavior signals such as textual search queries alongside watch histories across multiple recommender applications ({\em e.g.} directly recommended from homepage, suggested by features like what to watch next~\cite{zhao2019recommending}). This behavioral data includes temporal and engagement signals from user histories, such as interaction timestamps and likes or dislikes.

\paragraph{Video Representation.} Instead of randomly assigned video IDs (VIDs), we represent each video with a Semantic ID (SID)~\cite{rajput2023TIGER}: a hierarchical sequence of discrete codewords derived from the video's content features via a Residual-Quantized Variational AutoEncoder (RQ-VAE)~\cite{he2025plum}. SIDs offer two key advantages for user modeling: better generalization across head and tail videos~\cite{singh2024better}, and greater temporal stability, since VIDs suffer from vocabulary churn as the corpus evolves, which is a problem amplified when modeling long-range user histories.

\paragraph{Foundation Model Architecture.} While decoder-only models are popular in generative recommendation~\cite{he2025plum,zhang2025gpr,zhou2025onerecv2}, we adopt an encoder-decoder architecture for two reasons. First, encoders are known to capture more comprehensive sequential patterns on the full user history~\cite{ju2025generative}, where dense user embeddings can be naturally extracted from such contextualized representations, while the decoder autoregressively generates discrete SID-based user tokens. Second, the architecture offers deployment flexibility: the encoder and decoder can be decoupled for serving~\cite{zhou2025onerec,lyu2025dv365}, pairing a low-frequency heavy encoder for history compression with a high-frequency lightweight decoder for more recent behaviors. 

 During training, the decoder attends to the full encoder output via cross-attention and learns to generate SID tokens for targeted watches. At serving, dual-outputs are extracted: the dense user embedding is obtained via pooling ({\em e.g.}, last-token or mean pooling) over encoder outputs, and discrete user tokens are produced by applying beam search on the decoder to generate multiple SID sequences, each truncated to a coarse-grained prefix that serves as a single user token (details in Sec.~\ref{sec:enc_dec}).

\begin{figure}[t!]
    \centering
    \includegraphics[width=0.99\linewidth]{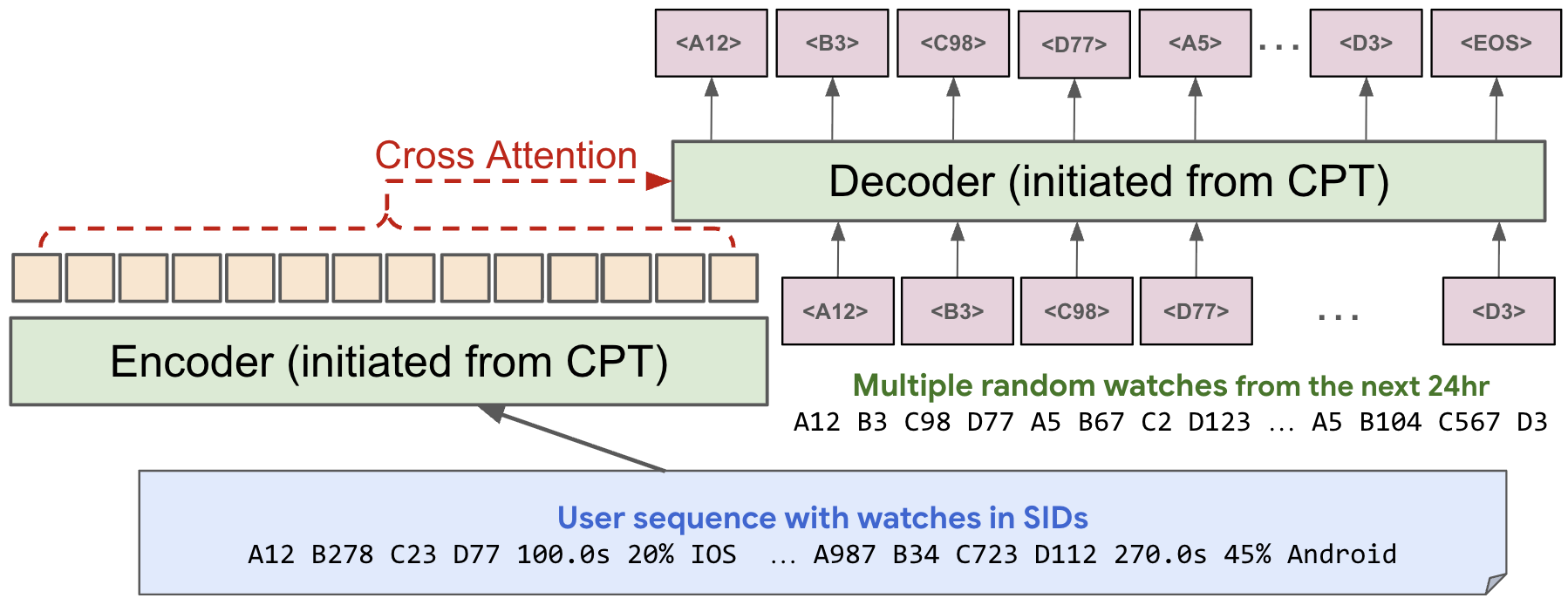}
    \caption{Training illustration for a single-scenario user example. The encoder takes a sequence of interleaved SID tokens and custom features. The decoder attends to the full encoder output via cross-attention and autoregressively generates SID tokens for multiple near-future target watches. Both are initialized from Continued Pre-Training (CPT)~\cite{he2025plum}.}
    \label{fig:enc_dec}
\end{figure}

\subsection{Model Training}
\label{sec:enc_dec}
Fig.~\ref{fig:enc_dec} illustrates training on a single-scenario user example, where chronological watches are encoded to predict multiple near-future watches. Specifically, the input sequence \texttt{A12 B278 C23 D77} represents the truncated prefix-4 of a video's hierarchical SID, while subsequent tokens such as \texttt{100.0s} and \texttt{IOS} encode numerical and textual features like watch time and device platform. Cross-scenario and search extensions appear in Sec.~\ref{sec:unified}.

\paragraph{Input Tokenization.}
Each watch $W_k$ is mapped via an $L_{\text{full}}$-level RQ-VAE to a hierarchical codeword sequence $[\text{SID}_{k,1}, \dots, \text{SID}_{k,L_{\text{full}}}]$; we retain only the coarse prefix-$L$ ($L{<}L_{\text{full}}$) to encourage diversity and mitigate memorization~\cite{zhou2025onerec} (ablated in Sec.~\ref{sec:exp_offline}). In the input sequence, each watch $W_k$ combines its prefix-$L$ SID as hard tokens with non-SID features encoded as either hard or soft tokens. Hard tokens are obtained by bucketing dense features or mapping categorical/text features to a vocabulary. Soft tokens are produced by embedding each feature independently, concatenating
the resulting representations, and projecting the concatenation to $M$
embeddings via an MLP.

\paragraph{Training Objective.}
Given a user's chronological watch sequence $W_1, \dots, W_n$, a cutoff timestamp $T$ partitions the sequence into a \textit{history} $[W_1, \dots, W_t]$ (watches before $T$) and a \textit{future window} $\{W_{t+1}, \dots, W_n\}$ (watches in $[T, T{+}24\text{h}]$). We adopt two key departures from standard next-watch prediction, both ablated in Sec.~\ref{sec:exp_offline}. First, we employ look-ahead sampling~\cite{xu2025redrec}: rather than predicting the immediate next watch $W_{t+1}$, we sample up to $N$ target watches $\{\tilde{W}_1, \dots, \tilde{W}_N\}$ from $\{W_{t+1}, \dots, W_n\}$, preventing overfitting to immediate continuations and capturing broader near-future interests. Second, predicting all $N$ targets jointly in a single decoder pass improves training efficiency over single-target supervision. Formally, denoting the prefix-$L$ SID tokens of target $\tilde{W}_i$ as $[\widetilde{\text{SID}}_{i,1}, \dots, \widetilde{\text{SID}}_{i,L}]$, \name{} minimizes:
\begin{equation}
\mathcal{L} = -\sum_{i=1}^{N} r(\tilde{W}_i) \cdot \sum_{j=1}^{L} \log P\!\left(\widetilde{\text{SID}}_{i,j} \;\middle|\; W_1, \dots, W_t, \tilde{W}_{<i}, \widetilde{\text{SID}}_{i,<j}\right).
\label{eq:loss}
\end{equation}
Here, $i$ indexes the $N$ sampled targets and $\mathcal{L}$ is computed only over the prefix-$L$ SID tokens. The engagement reward $r(\tilde{W}_i)$, combining multiple user satisfaction signals, encourages diverse and high-value consumption. In practice, rather than applying varying per-target weights in Eq.~\eqref{eq:loss}, we sample targets proportionally to $r(\tilde{W}_i)$~\cite{he2025plum} and weight the sampled targets equally ($r(\tilde{W}_i)=1$).

\subsection{Cross-Scenario and Search Integration}
\label{sec:unified}
The architecture described so far is trained and served separately for each content scenario. Consolidating long-form (LFV) and short-form (SFV) videos into a single model must account for their key differences: SFVs are browsed continuously without explicit click initiation~\cite{liu2023multitask}, resulting in a stronger feedback loop than LFVs, yet user interests are not isolated by format as nearly half of all users engage with both. Unifying these behaviors, together with textual user signals, is enabled by two forms of vocabulary sharing in \name{}. First, although LFVs and SFVs have disjoint video-ID spaces, both are encoded by the same RQ-VAE into one shared \emph{SID vocabulary}, in which their semantics overlap substantially (approximately 40\% at the first two prefix levels). Second, because the SID vocabulary extends the \emph{text vocabulary inherited from the pretrained LLM} and is grounded in it through CPT (Sec.~\ref{sec:overview}), textual signals such as search queries reside in the same token space as SID-based watch tokens and can be interleaved without a separate text encoder. Together, these properties allow a single model to improve computational efficiency, facilitate cross-scenario transfer, and capture a more comprehensive view of user preferences.

\begin{figure}[t!]
    \centering
    \includegraphics[width=0.99\linewidth]{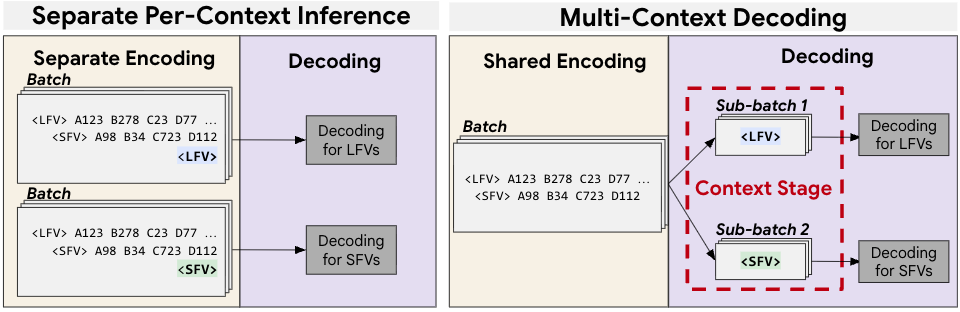}
    \caption{Comparison between separate per-context inference (left) and multi-context decoding (right). Multi-context decoding introduces a \textit{context stage} that creates parallel \textit{sub-batches} from a single shared encoder pass, enabling concurrent scenario-specific decoding with full reuse of encoder hidden states.}
    \label{fig:unified_decoding}
\end{figure}

\paragraph{Unified Training.}
Building on this unified token space, we construct a single input sequence that interleaves cross-scenario watches and textual search queries. Specifically, we prepend discrete scenario condition tokens (\texttt{<LFV>}, \texttt{<SFV>}) before each watch to distinguish content formats, and prepend a \texttt{<Search>} token to each of the $S$ most recent search queries, interleaving them chronologically within the watch sequence (ablated in Sec.~\ref{sec:exp_offline}). For the training objective, we extend the future target sampling (Sec.~\ref{sec:enc_dec}) to draw target watches across both SFVs and LFVs. Condition tokens are excluded from the loss, as predicting them is trivially easy and degrades SID prediction quality when included.

\paragraph{Multi-Context Decoding.} At serving time, a unified model must produce scenario-specific user tokens for each target context. A naive approach requires separate inference runs per context, redundantly processing the user history and negating the efficiency gains of a unified model. We propose \textit{multi-context decoding} to eliminate this redundancy. Given a shared encoder pass over the user history, we introduce a context stage that branches the decoding into parallel sub-batches, one per target context. Each sub-batch is initialized with its respective condition token and decoded independently via beam search, while all sub-batches fully share the encoder hidden states from a single encoder pass. As illustrated in Fig.~\ref{fig:unified_decoding}, this enables a single prefill pass with concurrent, context-specific decoding: generating LFV and SFV user tokens simultaneously from the same cached user representation.

\subsection{Downstream Adaptation}
\label{sec:integration}
Integrating discrete SID-based user tokens into downstream models, particularly LEMs that consume dense vectors, requires projecting each token into a continuous embedding. Drawing on item-SID taxonomies that distinguish static content reconstruction from learnable collaborative identity~\cite{li2026tokens,singh2024better}, we compare two token-to-embedding strategies:
(1)~\textbf{Prefix Embedding Mapping (EM)} reconstructs a \emph{static} content representation by mapping each predicted prefix-$L$ SID to the mean content embedding of all videos sharing that prefix; and
(2)~\textbf{Learnable Embedding (LE)} treats predicted prefixes as collaborative identity tokens, segmenting each SID into sub-words whose randomly initialized embeddings are summed and trained end-to-end with the downstream model (compared in Sec.~\ref{sec:exp_live}). Sub-words can be fixed-length $N$-grams of codewords ({\em e.g.}, unigram or nested hierarchical prefixes), or variable-length SentencePiece (SPM) pieces learned from item SID distributions~\cite{kudo2018spm}.

At serving, beam search yields $B$ SID sequences per user, each representing a predicted future interest. Downstream rankers aggregate these token embeddings into a compact representation in one of two ways before feeding it alongside other input features into the ranker's shared bottom or feature-interaction layers: (1) \emph{User-side pooling} condenses the $B$ embeddings into a single user-level representation (via mean, max, or top-$k$ pooling) with negligible serving overhead; (2) \emph{Candidate-aware target attention} uses the candidate item's SID as a query to attend over the $B$ user tokens, producing a candidate-specific representation for fine-grained matching.

Together, \name{}'s dual outputs offer different trade-offs in representation granularity and downstream integration overhead:
\begin{itemize}
    \item \textbf{User Embedding} (a single dense vector): Captures a holistic user state with negligible downstream training and serving overhead, serving as a drop-in feature for both two-tower ANN retrieval and ranking models.
    \item \textbf{User Tokens} ($B$ discrete SID prefixes): Preserve distinct future interests grounded in item semantics that a single pooled vector would otherwise average away, consumed by rankers through either user-side pooling or candidate-aware target attention.
\end{itemize}

\subsection{Serving System}
\label{sec:infra}

\begin{figure}[t!]
    \centering
    \includegraphics[width=0.9\linewidth]{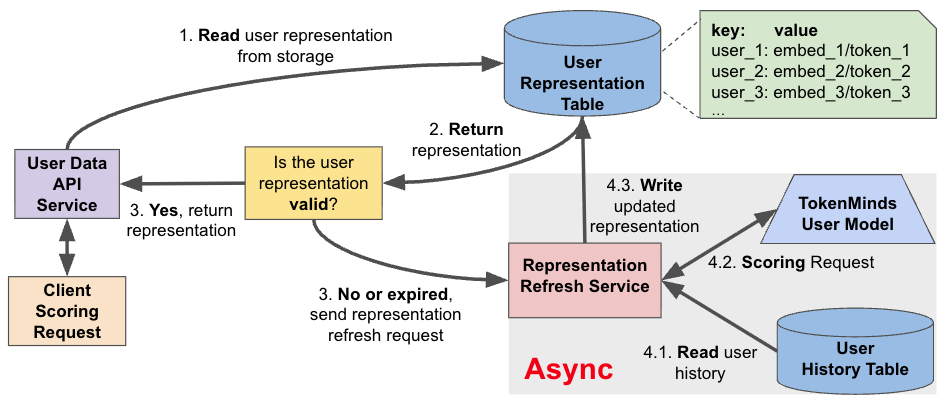}
    \caption{Asynchronous serving architecture of \name{}: real-time scoring retrieves cached representations directly; background representation refresh on cache miss or expiration (Steps 4.1--4.3).}
    \label{fig:infra}
\end{figure}

To serve billions of users under strict latency budgets, we decouple heavy generative inference from real-time scoring atop the User Behavior Service (UBS) framework~\cite{li2024ubs}. Fig.~\ref{fig:infra} illustrates the process: \name{} pre-computes user embeddings and discrete tokens asynchronously and stores them in a low-latency key-value cache (`User Representation Table'), keeping online scoring latency and cost independent of upstream \name{}'s complexity. When a scoring request arrives, the User Data API Service checks for valid, unexpired representations; if available, they are fetched and fed directly for downstream scoring (Steps 1--3). Otherwise, a background refresh is triggered immediately: the Representation Refresh Service reads the user's latest watch history, runs inference through the exported \name{} model, and writes the updated representations back to the cache (Steps 4.1--4.3).

\section{Experiments}
\label{sec:exp}
While \name{} has been deployed across YouTube's ranking, retrieval, and LLM-based production systems, we focus on ranking integration in this paper. We first detail the experimental setup and evaluation protocols (Sec.~\ref{sec:exp_setup}), followed by live A/B evaluations of \name{} across downstream production models (Sec.~\ref{sec:exp_live}), and finally the ablations and scaling analyses (Sec.~\ref{sec:exp_offline}). We investigate three research questions:

\textbf{RQ1 (Token Adaptation \& Viability):} How can discrete SID-based user tokens be effectively adapted for downstream continuous rankers, and do they constitute an effective user representation across industrial recommendation surfaces?

\textbf{RQ2 (Dual-Output Gains \& Comparison with LEM):} How do continuous user embeddings and discrete user tokens function together across production rankers, and how does the dual-output representation compare with the mature production LEM user model?

\textbf{RQ3 (Cross-Scenario Modeling Impact):} Can a unified, cross-scenario modeling paradigm (jointly training LFVs and SFVs) achieve computational efficiency gains without compromising downstream recommendation quality?

\addtocounter{table}{2} %
\begin{table*}[t!]
\centering
\caption{Live A/B evaluation of dual-output integration (\texttt{+Emb,+Token}) and comparison with the production LEM user model (\texttt{LEM}) across rankers. All baselines incorporate watched-item SIDs. $^{\dagger}$Denotes upgrading \name{} to longer user history length via soft-token compression~\cite{chen2026tokenfactory}. \textbf{Bold}$^{*}$: statistical significance at the 95\% confidence level.}
\label{tab:dual_and_consolidation}
\small
\begin{tabular}{lllccc}
\toprule
\textbf{Surface} & \textbf{Prod.\ Baseline} & \textbf{Treatment Arm (All Changes)} & \textbf{User Watch Time} & \textbf{Satisfied Engagement} & \textbf{Engaged Users} \\
\midrule
Shortspage & No \name{} & \texttt{+Emb} & $+0.05\%$ & $\mathbf{+0.09\%}^{*}$ & $+0.01\%$ \\ %
 & No \name{} & \texttt{+Token} & $+0.03\%$ & $\mathbf{+0.08\%}^{*}$ & $+0.04\%$ \\ %
 & No \name{} & \texttt{+Emb,+Token} & $\mathbf{+0.08\%}^{*}$ & $\mathbf{+0.13\%}^{*}$ & $\mathbf{+0.11\%}^{*}$ \\ %
\midrule
Homepage & \texttt{+LEM,+Emb} & \texttt{+LEM,+Emb,\textbf{+Token}} & $\mathbf{+0.10\%}^{*}$ & $\mathbf{+0.11\%}^{*}$ & $+0.02\%$ \\ %
 & \texttt{+LEM,+Emb,+Token} & \texttt{\textbf{-LEM},+Emb$^{\dagger}$,+Token$^{\dagger}$} & $\mathbf{+0.04\%}^{*}$ & $\mathbf{+0.05\%}^{*}$ & $\mathbf{+0.02\%}^{*}$ \\ %
\midrule

Watchpage & \texttt{+LEM,+Emb} & \texttt{+LEM,+Emb,\textbf{+Token}} & $\mathbf{+0.03\%}^{*}$ & $\mathbf{+0.03\%}^{*}$ & $+0.00\%$ \\
 & \texttt{+LEM,+Emb} & \texttt{\textbf{-LEM},+Emb,\textbf{+Token}} & $+0.02\%$ & $+0.02\%$ & $-0.01\%$ \\
\bottomrule
\end{tabular}%
\vspace{1ex}
\parbox{\textwidth}{\scriptsize \textit{Note:} Reported metrics may represent proprietary variants of the listed descriptors for each application. Shortspage Row 2 is the same study as Shortspage \texttt{+Token} row in Table~\ref{tab:single_output} (Panel A). Shortspage Row 1 is different from Shortspage \texttt{+Emb} row in Table~\ref{tab:single_output} (Panel B); the latter represents a subsequent production launch on an evolved Shortspage baseline.}
\end{table*}
\addtocounter{table}{-3} %

\subsection{Experiment Set-up}
\label{sec:exp_setup}
\paragraph{Model \& Training.} \name{} adopts an encoder-decoder architecture based on Gemini V1.5, comprising a 370M-parameter Mixture-of-Experts (MoE) encoder and a 370M-parameter dense decoder. Both are initialized from Continued Pre-Training (CPT) checkpoints~\cite{he2025plum}. \name{} is trained on YouTube LFV and SFV watch histories interleaved with textual search queries, using an internal JAX~\cite{jax2018github} and Pathways~\cite{barham2022pathways} framework developed for the Gemini ecosystem. All experiments use internal logs per YouTube's privacy policies, representing users with latent embeddings and coarse SID prefixes without exposing raw identifiers. We adopt continuous training on the latest daily data to capture fresh engagement signals, processing hundreds of millions of examples per day, which is more sample-efficient than traditional LEMs that typically require billions of interactions. Constrained by a fixed input token budget, we interleave the most recent few thousand watch interactions ($\mathcal{O}(1\text{K})$) with $S{=}50$ search queries (Sec.~\ref{sec:unified}). Each watch initially consists of a condition token, prefix length $L=4$ SID tokens (truncated from $L_{\text{full}}{=}8$ RQ-VAE levels with $2{,}048$ codewords per level), and $M{=}1$ soft token for non-SID features (Sec.~\ref{sec:enc_dec}), while each search query consists of a \texttt{<Search>} token followed by its text tokens; the upgraded
version ($\dagger$ in Table~\ref{tab:dual_and_consolidation}) compresses each watch or search query into a single soft token~\cite{chen2026tokenfactory}, extending the covered history several-fold within the same budget.
We sample prediction targets up to $N{=}15$ watches from the 24-hour look-ahead window and grid search for learning rates (detailed in Appx.~\ref{app:hyper}).
 
\paragraph{Serving at Scale.}
\name{} is served via the asynchronous processing infrastructure (Sec.~\ref{sec:infra}) with a 24-hour refresh cadence. For each user, we extract a 1152-dimensional dense embedding from the encoder and decode a total of $B{=}40$ SID sequences covering both LFV and SFV via beam search, each truncated to prefix length $L=4$ to form the user tokens. Joint token and embedding generation requires approximately $339\text{ms}$ per user, fully absorbed in background processing. Each user's discrete token representation requires only $1280$~bytes compared to $4608$~bytes for the dense embedding---a $72\%$ storage reduction.
At serving time, the system achieves a $96.4\%$ cache hit rate across $1.44\text{M}$ read requests per second from multiple production surfaces.

\paragraph{Online Evaluation \& Metrics.}
We evaluate the real-world impact of \name{} through live A/B experiments, integrating user embeddings and tokens across 3 primary production surfaces: Homepage (LFV), Watchpage (LFV) and Shortspage (SFV). Experiments were conducted for 10–14 days in general. We report performance across 3 key quality metrics~\cite{wang2026genie,li2026tokens,liu2023multitask}. \emph{User Watch Time} (User WT): total or time-capped platform watch duration. \emph{Satisfied Engagement} (Satisfied Eng.): value-weighted engagement capturing post-watch user satisfaction. \emph{Engaged Users} (Eng. Users): daily active users reaching engagement thresholds.

\paragraph{Production baselines \& LEM user model.}
Production rankers already consume SIDs of the user's watched videos~\cite{singh2024better}, and dense user embeddings from a mature non-LLM sequential user model (LEM). This model is a causal Transformer trained over user watch histories on next-item and engagement objectives. It has $\mathcal{O}(100\text{M})$ dense parameters and $\mathcal{O}(10\text{B})$ sparse video-related ID embedding parameters and is trained on $\mathcal{O}(1\text{B})$ user examples/day, roughly $20\times$ \name{}'s daily volume. All reported gains are therefore measured on top of both signals.

\subsection{Online Results}
\label{sec:exp_live}

\begin{table}[t!]
\caption{Live pivot study (110M model, Shortspage) comparing token adaptation strategies on surface-scoped SFV metrics: Prefix Embedding Mapping (EM) vs.\ Learnable Embedding (LE). \textbf{Bold}$^{*}$: statistical significance at the 95\% confidence level.}
\label{tab:token_adaptation}
\centering
\small
\begin{tabular}{@{}lccc@{}}
\toprule
\textbf{Strategy}  & \textbf{User WT}& \textbf{Satisfied Eng.} & \textbf{ Eng. Users} \\
\midrule

 EM (frozen codebook) & $-0.07\%$ & $-0.02\%$ & $+0.07\%$ \\
LE (co-trained table) & $+0.14\%$ & $\mathbf{+0.22\%}^{*}$ & $+0.08\%$ \\
\bottomrule
\end{tabular}%
\vspace{1ex} %
\parbox{\columnwidth}{\scriptsize \textit{Note:} Unlike subsequent tables that report platform-wide metrics, this table reports surface-scoped SFV metrics to maximize sensitivity for the lightweight 110M pivot study.}
\end{table}

\paragraph{Token Adaptation Pivot Study (RQ1).} Before committing massive compute to the full-scale dual-output architecture, we conducted an online pivot study using a lightweight 110M model on Shortspage to determine the optimal token adaptation strategy. As shown in Table~\ref{tab:token_adaptation}, we compared
frozen codebook based Prefix Embedding Mapping (EM) against downstream co-trained Learnable Embeddings
(LE). Specifically, for LE we used unigram tokenization to match existing item SIDs on Shortspage. 
We see that while EM is neutral across all key metrics, LE achieves a statistically significant gain in Satisfied Engagement. Auxiliary evaluations on Watchpage (SPM-based) and Homepage (nested $N$-gram-based) exhibited the same trend, confirming that learning a downstream task-specific embedding space outperforms static content mapping~\cite{singh2024better}.
 To verify that \texttt{LE}'s gains stem from \name{}'s predicted user tokens rather than from the extra parameters of the learnable embedding table, we ablate the shared SPM SID embedding table on Watchpage w/ and w/o \name{} user tokens (Appx.~\ref{app:param_control}): adding the table indexed only by context and candidate SIDs yields neutral CTR/AUC ($-0.002\text{\%pt}$ vs.\ baseline at convergence), whereas also indexing that same table with \name{} user tokens improves CTR/AUC by $+0.074\text{\%pt}$.  We thus utilize LE exclusively for all subsequent full-scale evaluations.

\begin{table}[t!]
\centering
\caption{Live A/B evaluation of integrating user tokens alone (\texttt{+Token}) and embeddings alone (\texttt{+Emb}) into production rankers. Production baselines: w/o \name{} features; w/ watched-item SIDs and LEM user embeddings. \textbf{Bold}$^{*}$: statistical significance at the 95\% confidence level.}
\label{tab:single_output}
\resizebox{\columnwidth}{!}{%
\begin{tabular}{lcccc}
\toprule
\textbf{Surface} & \textbf{Representation} & \textbf{User WT} & \textbf{Satisfied Eng.} & \textbf{Eng. Users} \\
\midrule
\multicolumn{5}{l}{\textit{Panel A: User Token Integration Studies (\texttt{+Token} vs.\ No-\name{} Baseline)}} \\
Homepage & \texttt{+Token} & $\mathbf{+0.13\%}^{*}$ & $+0.07\%$ & $\mathbf{+0.06\%}^{*}$ \\ %
Watchpage & \texttt{+Token} & $\mathbf{+0.06\%}^{*}$ & $\mathbf{+0.07\%}^{*}$ & $\mathbf{+0.04\%}^{*}$ \\ %
Shortspage & \texttt{+Token} & $+0.03\%$ & $\mathbf{+0.08\%}^{*}$ & $+0.04\%$ \\ %
\midrule
\multicolumn{5}{l}{\textit{Panel B: User Embedding Launches (\texttt{+Emb} vs.\ No-\name{} Baseline)}} \\
Homepage & \texttt{+Emb} & $\mathbf{+0.07\%}^{*}$ & $\mathbf{+0.09\%}^{*}$ & $\mathbf{+0.04\%}^{*}$ \\ %
Watchpage & \texttt{+Emb} & $\mathbf{+0.04\%}^{*}$ & $\mathbf{+0.04\%}^{*}$ & $\mathbf{+0.07\%}^{*}$ \\ %
Shortspage & \texttt{+Emb} & $\mathbf{+0.06\%}^{*}$ & $+0.05\%$ & $\mathbf{+0.10\%}^{*}$ \\ %
\bottomrule
\end{tabular}%
}
\vspace{1ex}
\parbox{\columnwidth}{\scriptsize \textit{Note:} Reported metrics may represent proprietary variants of the listed descriptors per surface.}
\end{table}

\paragraph{Single-Output Viability Across Surfaces (RQ1).}
We isolate the standalone impacts of discrete SID-based tokens (\texttt{+Token}) and dense embeddings (\texttt{+Emb}) across surfaces in Table~\ref{tab:single_output}. Panel A (\texttt{+Token}) demonstrates that SID-based user tokens independently add value to highly optimized production systems, directly answering the latter half of RQ1. Panel B shows the deployments of \texttt{+Emb} across surfaces. Beyond final-stage rankers, \texttt{+Emb} extends successfully into upstream candidate retrieval and mid-stage prescoring models, where latency budgets or two-tower ANN constraints make token usage difficult. For example, early-fusing \texttt{+Emb} into a sequential candidate retrieval model on Homepage and Watchpage improved live Satisfied Engagement by $\mathbf{+0.08\%^*}$.

\paragraph{Additive Dual-Output Gains (RQ2).} 
To evaluate the additive benefits of our dual-output integration, we introduce tokens alongside existing embeddings (\texttt{+Emb,+Token}) in production rankers. Table~\ref{tab:dual_and_consolidation} shows that this dual-output arm consistently performs best on Shortspage. Furthermore, on both Homepage (Row 1) and Watchpage (Row 1), adding \texttt{+Token} on top of strong production baselines already equipped with \texttt{+Emb} and \texttt{LEM} delivers further improvements in User Watch Time ($+0.10\%^{*}$ and $+0.03\%^{*}$) and Satisfied Engagement ($+0.11\%^{*}$ and $+0.03\%^{*}$). This confirms the additive dual-output gains, closing the first half of RQ2.

\paragraph{Comparison with the LEM User Model (RQ2).}
We next compare \name{}'s dual representation with the production LEM user model (Sec.~\ref{sec:exp_setup}). Across Tables~\ref{tab:single_output} and~\ref{tab:dual_and_consolidation}, the results reveal a clear asymmetry: adding \name{} on top of rankers that already consume LEM yields statistically significant gains across all surfaces, whereas once \name{}'s dual outputs are present, LEM adds no measurable residual value. In Table~\ref{tab:dual_and_consolidation} (Row~2 of Homepage/Watchpage), removing LEM alongside \name{}'s dual outputs still achieves statistically significant gains on Homepage (with soft-token compression~\cite{chen2026tokenfactory}) and maintains full parity on Watchpage. Direct comparisons against concurrent arms that retain LEM alongside the same dual outputs confirm that keeping LEM adds no measurable value on either Homepage ($-0.02\%$ [$-0.11$, $+0.07$] User WT, $-0.01\%$ [$-0.10$, $+0.07$] Satisfied Eng., $+0.01\%$ [$-0.04$, $+0.06$] Eng.\ Users; 95\% CIs) or Watchpage ($+0.01\%$ [$-0.02$, $+0.03$] User WT, $+0.01\%$ [$-0.02$, $+0.04$] Satisfied Eng., $+0.01\%$ [$-0.02$, $+0.04$] Eng.\ Users; 95\% CIs). Notably, an ablation on Watchpage that removes LEM while deploying only the dense embedding (w/o user tokens) degrades engagement ($-0.03\%$ User WT, $\mathbf{-0.04\%}^{*}$ Satisfied Eng., $-0.02\%$ Eng.\ Users), showing that the dense embedding alone is insufficient and that user tokens are needed to subsume LEM. Thus, \name{}'s dual outputs capture the signal LEM contributes and more, enabling LEM deprecation on both rankers at negligible serving TPU cost, answering the second half of RQ2.

\addtocounter{table}{1}
\begin{table}[t!]
\caption{Downstream integration overhead (Shortspage). Live request latency: mean
end-to-end server-side latency of the ranking requests in the live A/B. Serving throughput:
model server's max QPS at 85\% TPU duty cycle in offline load tests.}
\label{tab:sfv_cost}
\centering
\small
\setlength{\tabcolsep}{4pt} 
\begin{tabular}{@{}lcc@{}}
\toprule
\textbf{Metric} & \texttt{+Token} & \texttt{(+Emb,+Token)} \\
\midrule

Training Speed &  $-0.7\%$ &$-4.2\%$ \\ %
Live Request Latency (end-to-end) & $+0.08\%$ & $+0.17\%$ \\ %
Serving Throughput (max QPS)  & $-1.3\%$ & $-7.4\%$ \\ %
\bottomrule
\end{tabular}
\end{table}

\paragraph{Downstream Integration Overhead.}
Because \name{} pre-computes user representations asynchronously off the critical request path (Sec.~\ref{sec:infra}), downstream overhead is governed solely by how the ranker consumes each output format. Consuming the dense embedding alone (\texttt{+Emb}) is resource-neutral on every surface: the Shortspage launch measured $-0.3\%$ training speed and $-0.07\%$ live latency (within measurement noise), while the Watchpage ($0\%$ training speed, $\leq 0.01\%$ live latency) and Homepage ($\leq 0.01\%$ live latency) embedding launches were likewise neutral. Discrete tokens introduce a co-trained embedding table and aggregation compute that depend on the adaptation mechanism (Sec.~\ref{sec:integration}). With \emph{user-side pooling} on Homepage, where $B{=}40$ token embeddings (via nested $N$-gram hashing) are mean-pooled and fed alongside other input features, \texttt{+Token} incurs $-1.2\%$ in training speed with no measurable serving impact, as it avoids per-candidate attention. With \emph{candidate-aware target attention} on Shortspage, Table~\ref{tab:sfv_cost} quantifies the token arms of a controlled study over $B{=}40$ tokens: \texttt{+Token} costs $-0.7\%$ training speed and $-1.3\%$ serving throughput, and ingesting the dual outputs \texttt{(+Emb,+Token)} costs $-4.2\%$ training speed and $-7.4\%$ serving throughput, while end-to-end live request latency rises by at most $+0.17\%$, which in practice is negligible.

\begin{table}[t!]
\centering
\small
\setlength{\tabcolsep}{3.8pt}
\caption{Online evaluation of cross-scenario modeling on SFV and LFV rankers (Panel A) and upstream efficiency (Panel B). \textbf{Bold}$^{*}$: statistical significance at the 95\% confidence level.}
\label{tab:unified_results}
\begin{tabular}{@{}lcccc@{}}
\toprule
\multicolumn{5}{@{}l@{}}{\textit{Panel A: Recommendation Quality (vs.\ LFV-Only \name{})}} \\
\addlinespace[2pt]
\textbf{Surface} & \textbf{User WT} & \textbf{Satisfied Eng.} & \textbf{Eng. Users} & \textbf{Fresh Eng.} \\
\cmidrule{1-5}
Shortspage (SFV) & $+0.02\%$ & $+0.03\%$ & $+0.02\%$ & $\mathbf{+0.33\%}^{*}$ \\
Homepage (LFV)   & $+0.01\%$ & $+0.03\%$ & $+0.00\%$ & $\mathbf{+0.19\%}^{*}$ \\
\midrule
\multicolumn{5}{@{}l@{}}{\textit{Panel B: Upstream Compute (vs.\ 2 Separate Models)}} \\
\addlinespace[2pt]
\multicolumn{2}{@{}l}{\textbf{Training:} $-50\%$ Compute} & & \multicolumn{2}{r@{}}{\textbf{Serving:} $-31\%$ Compute} \\
\bottomrule
\end{tabular}
\end{table}
\paragraph{Cross-Scenario Modeling (RQ3).}
We report the quality and efficiency impacts of the unified cross-scenario model in Table~\ref{tab:unified_results}, across LFV surface Homepage and SFV surface Shortspage. Because initial production deployments across both Homepage and Shortspage consumed representations from the deployed LFV-trained model, online A/B quality (Panel~A) is evaluated against this production LFV-only baseline, whereas upstream resource savings (Panel~B) are measured against the counterfactual cost of maintaining two separate scenario-specific models. Panel A shows that replacing the LFV-only baseline with
unified representations consistently improves Freshness Engagement (\emph{Fresh Eng.}, measuring user interaction with new videos) across SFV and LFV surfaces, without degrading core engagement metrics. Notably, the unified model processes both LFV
and SFV watches within the same fixed-length input sequence, almost
halving the available LFV history. The fact that core metrics remain
neutral despite this reduction suggests that cross-scenario signals from
SFV interactions compensate for the reduced LFV context. Panel B shows that consolidating LFV and
SFV into a single framework yields substantial upstream cost savings. Upstream training compute (daily continuous-training TPU allocation) is reduced by 50\%, as a single unified pipeline replaces two separate scenario-specific models of identical capacity. Upstream serving
cost is reduced by 31\%, as multi-context decoding
(Sec.~\ref{sec:unified}) shares a single encoder pass across
scenarios (481 chips) instead of running two full model passes (698 chips
total). Downstream integration cost remains neutral, since the unified
model produces outputs of identical shape to the LFV-only baseline. Together, these results answer RQ3, confirming that the unified paradigm achieves substantial efficiency gains without sacrificing quality.

\subsection{Ablations \& Scaling Studies}
\label{sec:exp_offline}

\begin{table}[t!]
\caption{Ablation study on training objectives. Recall 
evaluated on SID prefix-$L$ predictions.}
\label{tab:offline_results}
\centering
\small
\setlength{\tabcolsep}{4pt} 
\begin{tabular}{@{}lcc@{}}
\toprule
\textbf{Model} & \textbf{Session Recall@10} & \textbf{Cold-Start Recall@10} \\
\midrule
\name{} (Ours) & 0.291 & 0.210 \\
\midrule
w/o Multiple Targets & 0.265 (\textit{-8.9\%}) & 0.203 (\textit{-3.3\%}) \\
w/o Look-ahead Window & 0.278 (\textit{-4.5\%}) & 0.189 (\textit{-10.0\%}) \\
w/o SID Truncation & 0.247 (\textit{-15.1\%}) & 0.174 (\textit{-17.1\%}) \\
\bottomrule
\end{tabular}
\end{table}

\paragraph{Training Objective Ablations.}
We evaluate offline token accuracy with two metrics: \textit{Session Recall} uses the near-complete history $[W_1,\dots,W_{n-1}]$ as input and predicts the final watch $W_n$; and \textit{Cold-Start Recall} uses the truncated history $[W_1,\dots,W_t]$ and predicts a randomly sampled watch from the future window $\{W_{t+1},\dots,W_n\}$. In both cases, we generate the top-10 SID sequences (prefix-$L$) and compute Recall@10. We train all models on identical data until convergence and report in Table~\ref{tab:offline_results} alongside three ablations, each removing one training innovation while keeping evaluation fixed: (1)~\textit{w/o Multiple Targets}: reduces sampled targets from $N{=}15$ to 1; (2)~\textit{w/o Look-ahead Window}: replaces look-ahead sampling with standard next-watch prediction (input $[W_1,\dots,W_t]$, target $W_{t+1}$); (3)~\textit{w/o SID Truncation}: trains and encodes inputs with full-length SIDs ($L_{full}$) instead of prefix-$L$; at evaluation, predictions are still compared at the prefix-$L$ granularity to ensure metric comparability. All three degrade accuracy, with look-ahead removal and full-length SIDs being particularly detrimental to cold-start performance ($-10\%$ and $-17\%$ relative, respectively), confirming that temporal look-ahead sampling and coarse-grained prefixes are essential for generalizing beyond the immediate session.

\paragraph{Encoder Attention.}
We compare bidirectional encoder self-attention against causal encoder masking at matched capacity with identical causal decoding, over matched two-week continuous-training windows. Bidirectional encoding improves Session Recall@10 by $+3.8\%$ (LFV) and $+8.8\%$ (SFV), supporting the rationale in Sec.~\ref{sec:overview} that bidirectional encoding captures more comprehensive sequential patterns over the full history than causal attention~\cite{ju2025generative}.

\paragraph{LLM Initialization \& Search Query Integration.}
We first explore different initialization strategies by comparing initialization from a Continued Pre-Trained (\texttt{CPT}) checkpoint~\cite{he2025plum} against two baselines: a general-purpose \texttt{Pre-Trained} Gemini checkpoint (without SID-specific grounding) and \texttt{Random} initialization. Relative to \texttt{Random} initialization, \texttt{Pre-Trained} Gemini improves $+3.3\%$ in Session Recall@10 and $+5.7\%$ in Cold-Start Recall@10, whereas initializing from the \texttt{CPT} checkpoint achieves substantially larger gains of $+5.3\%$ in Session Recall@10 and $+8.7\%$ in Cold-Start Recall@10. This confirms that the semantic grounding of SIDs during CPT~\cite{he2025plum} is beneficial for downstream task-specific fine-tuning, beyond the general sequence modeling capabilities of the base LLM alone.

Building on this \texttt{CPT}-initialized backbone with interleaved search queries ($S{=}50$ most recent queries), we evaluate the impact in a live A/B experiment on the Homepage ranker. We observe a $\mathbf{+0.04\%}^{*}$ gain in User WT, and $\mathbf{+0.05\%}^{*}$ in Eng. Users, confirming that textual search queries provide complementary intent signals beyond passive watch histories in production ranking.

\paragraph{Representation Quality \& Stability.}
Beyond future-watch recall, we verify that \name{}'s dual outputs remain diverse and temporally stable across asynchronous updates (detailed in Appx.~\ref{app:rep_quality}):
(1)~\emph{Token Diversity:} Across 5K users, the $B{=}40$ beam-decoded SID sequences closely match the ground-truth 24-hour watch distribution on both token collision and prefix duplication CDFs (Fig.~\ref{fig:diversity}), confirming diverse multi-interest coverage without beam collapse.
(2)~\emph{Checkpoint Stability:} Across consecutive daily checkpoints (${\approx}40\text{K}$ users with fixed history snapshots), outputs maintain $0.990\text{--}0.9999$ embedding cosine similarity and $0.88\text{--}0.89$ prefix-weighted Jaro-Winkler token similarity~\cite{winkler1990metric}, ensuring smooth consumption by downstream models without representation drift (see Appx.~\ref{app:rep_quality} for additional split-history self-retrieval and perturbation analyses).

\begin{figure}[t!]
    \centering
    \includegraphics[width=0.95\linewidth]{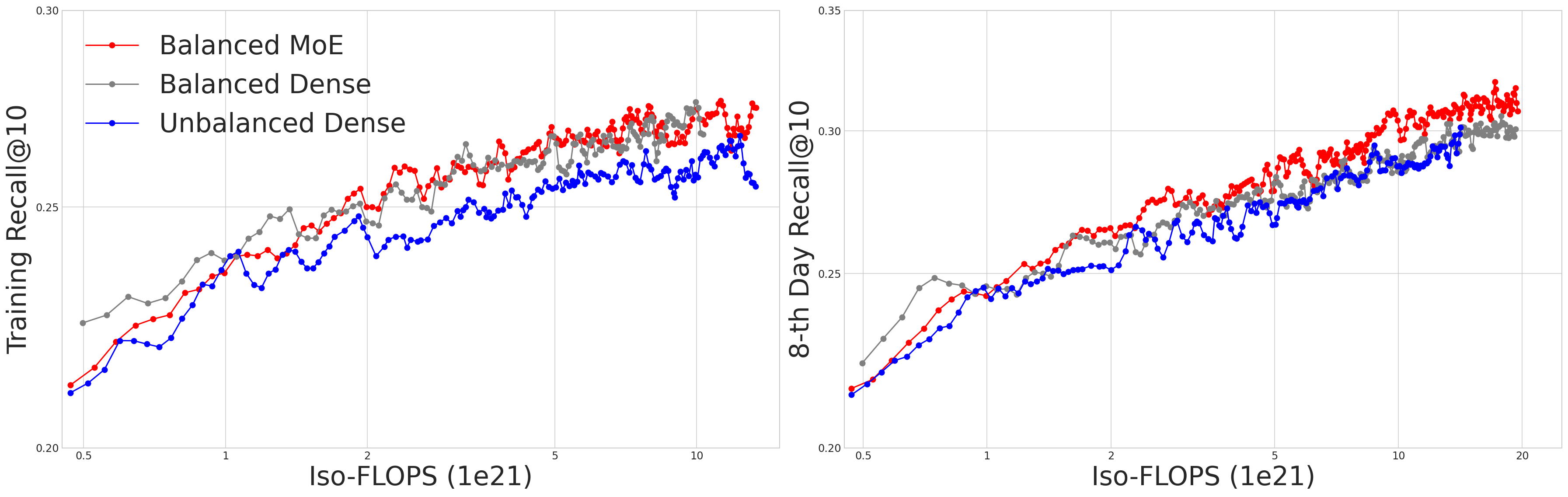}
    \caption{Training and 8th-Day Recall@10 against training Iso-FLOPS for model architecture variants for LFVs (SFVs omitted due to similar trends). All axes are in log-scale.
}
    \label{fig:model_variants}  
\end{figure}
\paragraph{Scaling Laws.}
To optimize computational efficiency, we adopt an accelerated offline protocol~\cite{he2025plum}, training on $7$ shuffled days and evaluating on the sequential $8$th day. 
We first compare three warm-started Gemini capacity variants against training Iso-FLOPS: \textit{Balanced MoE} (370M MoE Enc / 370M MoE Dec), \textit{Balanced Dense} (370M MoE Enc / 370M Dense Dec, our production architecture), and a purely dense \textit{Unbalanced Dense} variant (420M Dense Enc / 110M Dense Dec). Sweeping learning rates over $[0.1\times, 10\times]$ around the \textit{Balanced Dense} anchor, we observe two key trends from Fig.~\ref{fig:model_variants}: (1)~while \textit{Balanced MoE} achieves the highest 8th-Day Recall@10 at matched training FLOPS, we adopt \textit{Balanced Dense} for production because its dense decoder handles variable batch sizes and sequence lengths more efficiently during beam search, yielding higher serving throughput;
and (2)~\textit{Unbalanced Dense} matches \textit{Balanced Dense} on 8th-Day Recall despite a $3.4\times$ smaller decoder, validating the encoder-heavy capacity allocation discussed in Sec.~\ref{sec:overview}.

 Finally, scaling the training batch size from the 4K baseline to 8K and 16K (with independently tuned learning rates) improves SFV/LFV 8th-Day Recall@10 by $+2.5\%/{+}5.5\%$ and $+7.6\%/{+}13.7\%$, underscoring the benefit of maximizing batch size within hardware limits for compute-optimal training.

\section{Conclusion}
\label{sec:conclusion}
We introduced \name{}, a generative user modeling framework adapted from pre-trained LLMs with dual-outputs: \emph{continuous user embeddings} and \emph{discrete SID-based user tokens}. 
Cross-scenario modeling unifies LFV and SFV histories under a shared SID vocabulary with multi-context decoding, substantially reducing upstream
compute, while interleaving textual search queries captures explicit user intent alongside passive watch histories.
Deployed across YouTube's primary rankers since early 2026, \name{} demonstrates that discrete user tokens and dense embeddings provide additive gains across production models and outperform the non-LLM user model. Future work includes explicit objectives on the pooled encoder embedding and longer prediction horizons for lifelong histories.

\bibliographystyle{ACM-Reference-Format}
\bibliography{references}

\appendix
\appendix

\section{Additional Experimental Details}
\label{app:details}

\subsection{Hyper-parameter Tuning}
\label{app:hyper}
Through grid search over learning rates in $[10^{-6}, 10^{-3}]$, we find that linear warmup with cosine decay and cyclic restarts\cite{loshchilov2016sgdr,brown2020gpt3} achieves peak same-day performance, but a constant learning rate proves more robust to day-to-day distribution shifts under continuous training.

\begin{figure*}[t!]
    \centering
    \includegraphics[width=0.9\linewidth]{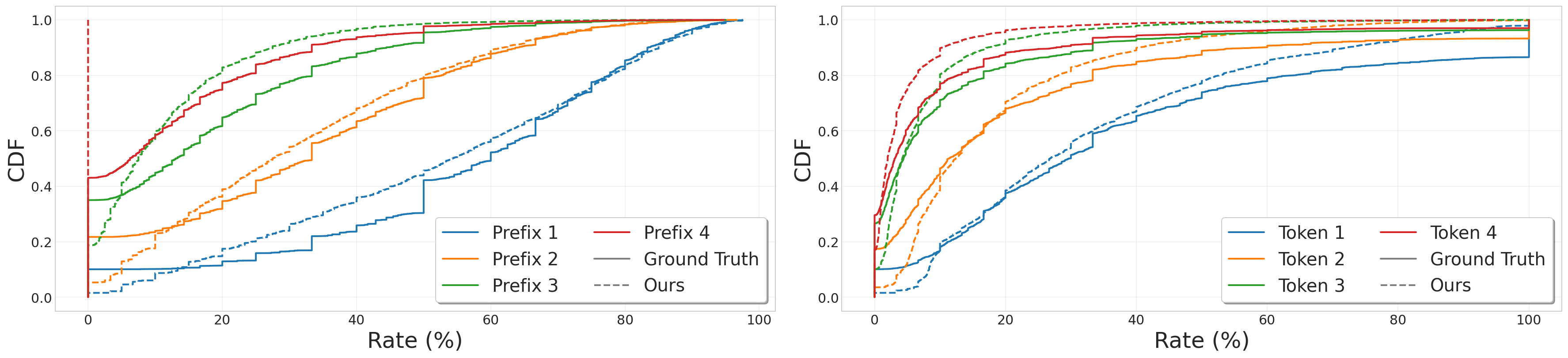}
    \caption{Diversity of generated user tokens vs. ground-truth user watches. CDFs show that beam search produces diverse predictions without collapsing into redundant outputs. (Left) SID Prefix Duplication Rate: fraction of identical prefixes. (Right) SID Token Collision Rate: fraction of identical tokens at each position.}
    \label{fig:diversity}
\end{figure*}

\subsection{Isolating User-Token Gains from Embedding-Table Capacity (Watchpage)}
\label{app:param_control}
A natural question when comparing Prefix Embedding Mapping (\texttt{EM}) and Learnable Embeddings (\texttt{LE}) in Sec.~\ref{sec:exp_live} is whether \texttt{LE}'s performance gains derive from the predictive signal in \name{}'s generated user tokens or merely from the additional parameters of the co-trained embedding table. On the Watchpage ranker, \texttt{LE} employs a shared SentencePiece Model (SPM)~\cite{kudo2018spm} SID embedding table with a vocabulary size of $8\text{M}$ and embedding dimension $256$, which embeds sub-word segments of context (current watch), candidate, and \name{} user SID tokens before candidate-aware cross-attention.

To isolate the contribution of \name{} user tokens from the parameter capacity of this $8\text{M} \times 256$ SPM table, we conducted a controlled three-arm offline ablation on the Watchpage ranker (with other dense user embeddings ablated to avoid feature shadowing), evaluated over steps $[1.35\text{M}, 1.45\text{M}]$ at convergence (Table~\ref{tab:app_param_control}):
\begin{enumerate}
    \item \textbf{Baseline (\texttt{base}):} Watchpage ranker without the shared SPM SID embedding table or \name{} user tokens.
    \item \textbf{Table-Capacity Control (\texttt{base\_rqvae}):} Adds the full $8\text{M} \times 256$ learnable SPM embedding table and cross-attention projections, indexed by context and candidate video SIDs, but \emph{without} \name{} user tokens.
    \item \textbf{\name{} User Tokens (\texttt{rqvae\_token}):} Indexes the exact same $8\text{M} \times 256$ SPM embedding table with both context/candidate SIDs and the $B{=}40$ \name{} user tokens.
\end{enumerate}
As shown in Table~\ref{tab:app_param_control}, adding the SPM embedding table and context/candidate SIDs alone (\texttt{base\_rqvae}) yields no improvement over \texttt{base} ($0.75387$ vs.\ $0.75389$ CTR/AUC, $-0.002\text{\%pt}$). By contrast, indexing the same table with \name{} user tokens (\texttt{rqvae\_token}) improves CTR/AUC to $0.75462$ ($+0.074\text{\%pt}$ over \texttt{base\_rqvae}), confirming that the downstream gain arises from \name{}'s generative user representations rather than extra embedding table capacity.

\begin{table}[t!]
\caption{Offline parameter-isolation control on the Watchpage ranker evaluating the impact of the shared SPM Learnable Embedding (\texttt{LE}) table with and without \name{} user tokens.}
\label{tab:app_param_control}
\centering
\resizebox{\columnwidth}{!}{%
\begin{tabular}{@{}lcccc@{}}
\toprule
\textbf{Arm} & \textbf{SPM Table} & \textbf{Ctx/Cand SIDs} & \textbf{User Tokens} & \textbf{CTR/AUC ($\Delta$)} \\
\midrule
\texttt{base} & No & No & No & $0.75389$ (---) \\
\texttt{base\_rqvae} & Yes & Yes & No & $0.75387$ ($-0.002\text{\%pt}$) \\
\texttt{rqvae\_token} & Yes & Yes & Yes & $\mathbf{0.75462}$ ($\mathbf{+0.074\text{\%pt}}$) \\
\bottomrule
\end{tabular}
}
\end{table}

\subsection{Representation quality}
\label{app:rep_quality}

\paragraph{Token Diversity.}
We verify that token accuracy is not achieved through
redundant beams. Since each decoded SID serves as a distinct interest signal for downstream models, beam collapse---where multiple beams converge to near-identical outputs---would severely limit representational capacity. We assess diversity using two metrics: \textit{SID Token Collision Rate} (measuring identical token frequency at position $x$, where high collision signifies semantic beam collapse) and \textit{SID Prefix Duplication Rate} (counting identical prefixes up to $x$, where high repetition indicates isolated exploration of a single branch). We evaluate these metrics over a sample of 5K users, comparing the generated tokens against ground-truth watches from the same look-ahead 24-hour window. To ensure a fair comparison despite size disparities, we downsample the larger set (ground-truth watches or generated SIDs) to match the size of the smaller set. To reduce the variance from subsampling, for each comparison, we repeat the downsampling of the larger set 10 times, and report the averaged similarity scores.
The resulting CDFs across different token indices, plotted in Fig.~\ref{fig:diversity}, demonstrate that \name{} achieves prediction diversity on par with the ground-truth distribution, confirming that beam search produces diverse interest signals without collapsing into redundant outputs.

\paragraph{Perturbation Stability.}
Complementary to our 40K-user production checkpoint stability monitoring (Sec.~\ref{sec:exp_setup}), we evaluate embedding consistency across 2K random sampled users. For each user, we generate an embedding $E_A$ from their full history and a perturbed variant $E_{A^*}$ by randomly dropping watches, then compare against a random user's embedding $E_B$. The average cosine similarity $\text{Sim}(E_A, E_{A^*}) = 0.993$ far exceeds $\text{Sim}(E_A, E_B) = 0.761$, indicating that embeddings are stable under minor history perturbations while remaining discriminative across users.

\paragraph{Split-History Self-Retrieval.}
Beyond random dropout perturbations mentioned above, we evaluate long-range user identity persistence across $1\text{K}$ users under two strict disjoint-history splits:
(1)~an \emph{even/odd split} (retrieving a user's odd-indexed watch embedding from their even-indexed watch embedding), where \name{} achieves $96.2\%\text{--}97.6\%$ Top-10 self-retrieval accuracy ($91.8\%\text{--}94.0\%$ Top-1; intra-class correlation $\text{ICC} = 0.858\text{--}0.892$); and
(2)~a \emph{chronological split} (retrieving a user's newer $50\%$ history embedding from their older $50\%$ history embedding), achieving $80.2\%\text{--}82.9\%$ Top-10 self-retrieval accuracy ($\text{ICC} = 0.671\text{--}0.741$) with local intrinsic dimensionality $\text{LID}_{k=15} \approx 10.9\text{--}11.5$.
This confirms that the encoder embedding captures persistent user preferences across non-overlapping interaction windows rather than merely memorizing recent watches.

\end{document}